\documentclass[dvips]{article}
\usepackage{icrctc07}

\title{Developments in Nanosecond Pulse Detection Methods \& Technology}
\shorttitle{Developments in Nanosecond Pulse Detection}

\authors{R. A. McFadden$^{1,2}$, N. D. R. Bhat$^{4}$, R. D. Ekers$^{2}$, C. W. James$^{3}$, D. Jones$^{3}$, S. J. Tingay$^{4}$, P. P. Roberts$^{2}$, C. J. Phillips$^{2}$, R. J. Protheroe$^{3}$}
\shortauthors{McFadden and Ekers et al.}
\afiliations{$^1$School of Physics, Univ. of Melbourne, VIC, Australia\\ $^2$Australia Telescope National Facility, Epping, NSW, Australia\\$^3$School of Chemistry \& Physics, Univ. of Adelaide, SA, Australia\\$^4$Centre for Astrophysics \& Supercomputing, Swinburne University of Technology, VIC, Australia}
\email{rmcfadde@physics.unimelb.edu.au}

\abstract{A promising method for the detection of UHE neutrinos is
the Lunar Cherenkov technique, which utilises Earth-based radio
telescopes to detect the coherent Cherenkov radiation emitted when a
UHE neutrino interacts in the outer layers of the Moon. The LUNASKA
project aims to overcome the technological limitations of past
experiments to utilise the next generation of radio telescopes in
the search for these elusive particles. To take advantage of
broad-bandwidth data from potentially thousands of antennas requires
advances in signal processing technology. Here we describe recent
developments in this field and their application in the search for
UHE neutrinos, from a preliminary experiment using the first stage
of an upgrade to the Australia Telescope Compact Array, to
possibilities for fully utilising the completed Square Kilometre
Array. We also explore a new real time technique for characterising ionospheric pulse dispersion which specifically measures ionospheric electron content that is line of sight to the moon.}

\begin{document}
\maketitle
\section{Introduction}

The origin of the most energetic particles observed in nature, the
ultra high energy (UHE) cosmic rays (CR), which have energies
extending up to at least $2 \times 10^{20}$ eV, is currently
unknown. Finding the origin of these particles will have important
astrophysical implications. However, direct detection of UHE
neutrinos is very difficult due to their extremely small interaction
cross-sections. Instead, they may be detected indirectly via
observation of the Askaryan effect \cite{askaryan62excess} in the
lunar regolith. Askaryan first predicted coherent Cherenkov emission
in dielectric media at radio and microwave frequencies. Using the
Moon as a large volume neutrino detector, coherent radio Cherenkov
emission from neutrino-induced cascades in the lunar regolith can be
observed with ground based telescopes. This method was first
proposed by Dagkesamanskii and Zheleznykh
\cite{dagkesamanskii89radio} and first applied by Hankins, Ekers and
O'Sullivan \cite{hankins96asearch} using the Parkes radio telescope.

Coherent Cherenkov radiation is a linearly polarised broadband
emission. The spectrum of coherent Cherenkov emission rises
approximately linearly with frequency until a peak value is reached.
The peak frequency is determined by de-coherence and/or attenuation
in the regolith, and can vary between a few hundred MHz and
approximately 5 GHz. The dependence of the peak frequency on shower
geometry makes the choice of an optimum observation frequency
non-trivial \cite{clancy07lunar}.

\section{Detection Issues}

Lunar Cherenkov emission produces an extremely narrow pulse
(sub-nanosecond duration). These pulses travel through the
ionosphere and experience a frequency dependent time delay resulting
in pulse dispersion similar to the dispersion experienced by pulsar
pulses traveling through the inter-stellar medium. Therefore lunar
Cherenkov pulses received on an Earth based radio telescope will be
dispersed by the ionosphere and further broadened by receiver band
limiting.

\begin{figure}
\begin{center}
\noindent 
\includegraphics [width=0.45\textwidth]{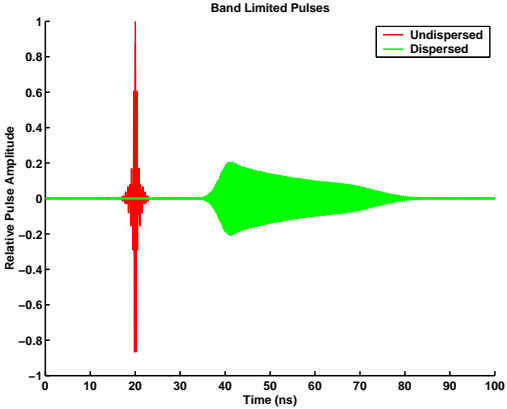}
\end{center}
\caption{Pulse dispersed over 50ns (corresponding to solar maximum)
and bandlimited between 1.2-1.8 GHz.}\label{fig1}
\end{figure}

The ionospheric pulse dispersion must be known in real time to
maximise the received signal to noise ratio and subsequent chances
of pulse detection. Coherent pulse de-dispersion requires an
accurate knowledge of the ionospheric dispersion characteristic
which can be parameterised by the instantaneous ionospheric Total
Electron Content (TEC). The TEC produces a frequency dependent delay
in the received signal via Equation \ref{eq:time_delay}

\begin{equation}
\Delta t=1.34\times10^{-7} \times TEC \times \nu^{-2}
\label{eq:time_delay}
\end{equation}

where $\Delta t$ is in seconds, $TEC$ in electrons per m$^{2}$
and frequency in Hz.

As Cherenkov radio pulses are much shorter in time than any signals
normally encountered in radio astronomy, real time de-dispersion and
detection requires broadband technology and innovations to the
current instrumentation. Our experiment utilises the Australia
Telescope Compact Array radio telescope
which currently has 600 MHz bandwidth at 1.5 GHz but is being
upgraded to a 2 GHz bandwidth over 1-3 GHz. Due to data storage limitations, the only
way to exploit these new bandwidths is to implement real time
detection algorithms. This requires innovations in hardware design
to perform signal processing with nanosecond timing accuracy. New
radio instruments planned, such as the Square Kilometre Array (SKA),
could provide huge advances in collecting area and technology for
the method of lunar Cherenkov UHE neutrino detection.


An array of small dishes is the optimum radio instrument for this
experiment. If the dish size is kept small enough ($\sim$20m at 1.4 GHz),
all of the Moon will be seen by each antenna's primary beam. Increased
sensitivity from using larger dishes is off-set by decreased
coverage of the Moon due to the smaller beam size, and thus a
reduced effective aperture. Using multiple smaller dishes allows the
same sensitivity without loss of coverage, provided their separation
is such that the thermal lunar emission --- which tends to dominate the
system temperature --- is incoherent between antenna. Signals can also be added coherently to form
multiple beams around the limb of the Moon, where the event rate is
maximised, and the array geometry can be exploited for RFI
discrimination based on the signal direction of arrival.

\section{Current Experiment with the ATCA}

Our current experiment uses the Australia Telescope Compact Array
with a 600 MHz bandwidth. The 600 MHz signal is taken from a
maintenance point in the antenna, therefore obtaining this bandwidth
requires the development of customised detection hardware which must
operate with nanosecond timing accuracy. Ionospheric pulse
dispersion must also be corrected in real time to maximise the
received signal to noise ratio for threshold detection
\cite{hankins96asearch}. For this stage of our experiment, we have
developed analog de-dispersion filters and a field-programmable gate
array (FPGA)-based trigger system to detect events in real time and
transmit event data to the site control room via an Ethernet link.

FPGAs are reprogrammable semiconductor devices, composed of Configurable Logic Blocks (CLBs), which can perform simple logic tasks as well as more complicated mathematical and signal processing algorithms. Their reconfigurable logic blocks also provide the ability to optimise hardware interconnections for each specific algorithm implemented. Modern FPGAs can operate at clock frequencies of around 500 MHz and have signal processing capabilities that far exceed anything previously available for optimised real time operation. This high speed operation and optimised processing power make FPGAs an ideal technology to perform real time signal processing and pulse detection.

The preliminary ATCA experiment makes use of an FPGA based sampler
board for event triggering. Future experiments with the ATCA will
make use
of larger FPGA boards to perform real time de-dispersion 
however for this stage of the experiment, signal de-dispersion is performed
in analog microwave filters with a fixed dispersion characteristic.
These filters were designed using a new method of planar microwave
filter design based on inverse scattering  \cite{roberts95design}. This results in
filters with a continuously changing profile, in this case a
microstrip line with continuously varying width. The width
modulations on the microstrip line produce cascades of reflections
which sum to produce the desired frequency response. For the lunar
Cherenkov detection experiment the desired frequency response has a
group delay which is quadratically chirped according to Equation \ref{eq:time_delay}. As the microwave de-dispersion
filters have a fixed de-dispersion characteristic, an estimate had
to be obtained for the TEC which would minimise the errors
introduced by temporal ionospheric fluctuations.

Fluctuations in the ionosphere experience a strong diurnal cycle and
are also dependent on the season of the year, phase of the current
(11-year) solar cycle and the geometric latitude of observation.
Observations were planned during night-time hours to minimize these
fluctuations and as we are currently at solar minimum, it was
assumed that the electron density for our observations could be
estimated based on measurements from the corresponding season last
year. This estimate was produced by analysis of Vertical TEC (VTEC)
data maps which were derived from GPS dual frequency signals and are
available online from NASA's Crustal Dynamics Data Information
System (CDDIS) \cite{CDDIS}.

Our observations were during the nights of May 5, 6 and 7 2007.
These dates were chosen to ensure that the Moon was at high
elevation (particularly during the night time hours of ionospheric
stability) and positioned such that we would be sensitive to UHE
particles from the galactic center. The corresponding CDDIS
measurements for the month of May 2006 gave an average VTEC of 7.06
TEC Units (1 TECU -- $10^{16}$ electrons per m$^2$) over night-time
hours (10pm-8am) with a standard deviation of 1.3 TECU. The filter
design assumed a differential delay of 5 ns across the 1.2-1.8 GHz
bandwidth, which was based on the average VTEC value, corrected for
slant angle through the ionosphere, and converted to a differential
time delay. GPS data available post experiment revealed that the
average VTEC for the nights of our observations was actually 7.01
TECU which gave an average differential delay of 4.39, with standard
deviation 1.52, corrected for slant angle.

Our event trigger has been implemented in an FPGA based sampling
board which was designed as part of the broadband upgrade planned for the compact array. The sampler board has
two inputs, sampled at 2.048 G samples/s, which we are using to
import two linear orthogonal polarisations of the 600 MHz RF signal.
The sampled inputs are then multiplexed into parallel 512 MHz data
streams for input into a Xilinx FPGA which performs pulse
detection. Detection is performed via thresholding on both
polarisation streams. The sampled data is buffered for 2 microseconds so
that data surrounding candidate events can be sent back and recorded
in the case of a detection.

Event data is transferred to a central processing site via a Gbit
Ethernet connection for off-line processing. The de-dipsersion
filters and CABB sampler boards were installed on three antennas so
that coincidence testing and direction of arrival discrimination could be
performed during off-line processing. As each antenna triggers
independently, trigger timing information and nanosecond
synchronisation is essential for both of these processing stages.

Our current sampler boards do not have the capability of time
stamping with nanosecond accuracy, and so alternative methods of
time stamping had to be investigated. As the system temperature is
dominated by thermal emission from the Moon, the array operates in the intensity interferometer regime \cite{brown57interferometry}, and there is a small amount
of correlation between the signal received at each antenna. This
level of correlation can be exploited to determine relative timing
between event buffers sent back from different antennas.

\section{Future Experiments and Developments in Pulse De-Dispersion}
\label{sec:Future}

Future improvements to the ATCA UHE neutrino detection experiment
include using 5 antenna, performing coherent signal combination in
real time to enable coincidence testing and an increase to 2 GHz
bandwidth. Real time coherent signal addition will be possible as
part of an upgrade planned for the ATCA which will include the
installation of powerful new FPGA based back-end receiver hardware.
This hardware can also be used to implement real time de-dispersion
algorithms and we have developed a technique for obtaining
measurements of the ionospheric TEC which are both instantaneous and
line-of-sight to the lunar observations. The ionospheric TEC can be
deduced from Faraday Rotation measurements of a polarised source
combined with geomagnetic field models, which are more stable than
ionospheric models \cite{ganguly01ionospheric}. The Faraday Rotation
induced in a radio wave is related to the ionospheric electron
content via Equation \ref{eq:Faraday_rotation}

\begin{equation}
\Omega = 2.36 \times 10^4 \nu^{-2}\int_{path}N(s)B(s)\cos \theta  ds
\label{eq:Faraday_rotation}
\end{equation}

where $\Omega$ is the rotation angle in radians, $\nu$ is the signal
frequency in Hz, $N$ is the electron density m$^{-3}$, $B$ is the
geomagnetic field strength in T, $\theta$ is the angle between the
direction of propagation and the magnetic field vector and ds is a
path element in m. Our novel approach is to use this technique with
the polarised thermal radio emission from the lunar limb as our
polarised source, to obtain instantaneous, line-of-sight TEC
measurements. This makes the lunar Cherenkov technique extremely
attractive for UHE cosmic ray and neutrino astronomy as it removes
the need for searching in dispersion space.

The next phase in our experiment involves the Square Kilometre Array
(SKA). The SKA is a new generation radio telescope array which will
be one hundred times more sensitive than the best present day
instruments. Current designs proposed for the SKA consist of large
numbers (~$10^4$) of small dishes (6-12m) to achieve a square km of
collecting area in the 0.1-3 GHz range which is critical for UHE
neutrino experiments. To gain the advantage of the large number of
small dishes offered by the proposed SKA designs, signals have to be
combined and analysed with nanosecond timing accuracy. This will
involve forming phased array beams in real time using special
purpose beam forming hardware. We will require enough beams to cover
the visible surface of the Moon over the entire frequency range, and
will thus have increasing resolution with frequency. Using these
methods, we expect the SKA sensitivity to reach the level at which a
flux of UHE neutrinos will be detected.

\section{Acknowledgements}
This research was supported under the Australian Research Council's
Discovery funding scheme (project number DP0559991). Professor R. D.
Ekers is the recipient of an Australian Research Council Federation
Fellowship (project number FF0345330)

\small
\bibliography{icrc0434}

\begin{thebibliography}{1}

\bibitem{askaryan62excess}
G.~A. {Askar'yan}.
\newblock {Excess negative charge of an electron-photon shower and its coherent
  radio emission}.
\newblock {\em Soviet Physics JETP-USSR}, 14(2):441--443, 1962.

\bibitem{brown57interferometry}
R.~H. {Brown} and R.~Q. {Twiss}.
\newblock {Interferometry of the Intensity Fluctuations in Light. I. Basic
  Theory: The Correlation between Photons in Coherent Beams of Radiation}.
\newblock {\em Royal Society of London Proceedings Series A}, 242:300--324,
  November 1957.

\bibitem{dagkesamanskii89radio}
R.~D. {Dagkesamanskii} and I.~M. {Zheleznykh}.
\newblock {A radio astronomy method of detecting neutrinos and other
  superhigh-energy elementary particles}.
\newblock {\em Pis ma Zhurnal Eksperimental noi i Teoreticheskoi Fiziki},
  50:233--235, September 1989.

\bibitem{ganguly01ionospheric}
S.~{Ganguly}, A.~{Brown}, A.~{DasGupta}, and S.~{Ray}.
\newblock {Ionospheric reconstruction using Faraday rotation data: A new
  technique}.
\newblock {\em Radio Science}, 36:789--800, 2001.

\bibitem{hankins96asearch}
T.~H. {Hankins}, R.~D. {Ekers}, and J.~D. {O'Sullivan}.
\newblock {A search for lunar radio Cherenkov emission from high-energy
  neutrinos}.
\newblock {\em MNRAS}, 283:1027--1030, December 1996.

\bibitem{clancy07lunar}
C.~W. {James}, R.~D. {Ekers}, R.~A. {McFadden}, and R.~J. {Protheroe}.
\newblock {The lunar Cherenkov technique: From Parkes onwards}.
\newblock In {\em Proceedings of XXX ICRC}, 2007.

\bibitem{CDDIS}
NASA.
\newblock Crustal dynamics data information system, 2007.
\newblock [Online; accessed 5-June-2007].

\bibitem{roberts95design}
P.~P. {Roberts} and G.~E. {Town}.
\newblock {Design of microwave filters by inverse scattering}.
\newblock {\em IEEE Trans. Microwave Theory and Techniques}, 43(4):739--743,
  1995.

\end{thebibliography}
\bibliographystyle{plain}
\end{document}